\documentclass[a4paper, prb, twocolumn, showpacs]{revtex4}

\usepackage{amssymb}
\usepackage{amsmath}
\usepackage{color}
\usepackage{graphics, graphicx}
\usepackage{psfrag}
\usepackage{mathcomp}
\usepackage{subfigure}

\newcommand{\ff}[1]{{\boldsymbol #1}}

\begin{document}

\author{Irakli Titvinidze}
\author{Andrej Schwabe}
\author{Niklas Rother}
\author{Michael Potthoff}

\affiliation{I. Institut f\"ur Theoretische Physik, Universit\"at Hamburg, Jungiusstra\ss{}e 9, 20355 Hamburg, Germany }


\title{Dynamical mean-field theory of indirect magnetic exchange}

\begin{abstract}

To analyze the physical properties arising from indirect magnetic exchange between several magnetic adatoms and between complex magnetic nanostructures on metallic surfaces, the real-space extension of dynamical mean-field theory (R-DMFT) appears attractive as it can be applied to systems of almost arbitrary geometry and complexity. 
While R-DMFT describes the Kondo effect of a single adatom exactly, indirect magnetic (RKKY) exchange is taken into account on an approximate level only.
Here, we consider a simplified model system consisting of two magnetic Hubbard sites (``adatoms'') hybridizing with a non-interacting tight-binding chain (``substrate surface''). 
This two-impurity Anderson model incorporates the competition between the Kondo effect and indirect exchange but is amenable to an exact numerical solution via the density-matrix renormalization group (DMRG).
The particle-hole symmetric model at half-filling and zero temperature is used to benchmark R-DMFT results for the magnetic coupling between the two adatoms and for the magnetic properties induced in the substrate.
In particular, the dependence of the local adatom and the nonlocal adatom-adatom static susceptibilities as well as the magnetic response of the substrate on the distance between the adatoms and on the strength of their coupling with the substrate is studied.
We find both, excellent agreement with the DMRG data even on subtle details of the competition between RKKY exchange and the Kondo effect but also complete failure of the R-DMFT, depending on the parameter regime considered.
R-DMFT calculations are performed using the Lanczos method as impurity solver.
With the real-space extension of the two-site DMFT, we also benchmark a simplified R-DMFT variant. 
\end{abstract}

\maketitle

\section{INTRODUCTION}
\label{intro}

The rapidly improving experimental techniques to probe magnetic adatoms on non-magnetic surfaces allow for direct studies of fundamental magnetic exchange mechanisms on an atomic scale.
Besides access to the structural and the electronic properties of such adatoms and of the underlying substrate for a given system, the construction of tailored magnetic model systems represents an exciting perspective. \cite{Eigler-Schweizer,LSBD98,HLH06,Wie09}  
Magnetic structures of nanometer size provide extremely small systems suitable to store and to transport information and may realize efficient nano spintronics devices. \cite{KCWW11}

The competition between an indirect magnetic exchange of the adatoms via the substrate electrons on the one hand and the screening of the adatom magnetic moment by the conduction-band electrons of the substrate on the other represents a prominent example for a physical problem becoming accessible to new real-space techniques.
The scanning tunneling microscope (STM) \cite{STM} has been used to investigate the Kondo physics \cite{Kondo} 
of single magnetic adatoms \cite{LSBD98,Madhavan,Wahl04} and the magnetic properties of the individual magnetic islands\cite{Yamasaki, Bode-Wiesendanger} on non-magnetic substrates. 
Using STM, it is possible to investigate the direct magnetic interaction of atom pairs. \cite{HLH06,Lee,Kitchen} 
Indirect magnetic exchange, i.e.\ the Ruderman-Kittel-Kasuya-Yosida (RKKY) interaction \cite{RKKY} between two adatoms, was detected through the Kondo effect. \cite{Wahl07}
A direct real-space study of the RKKY coupling, however, comes in reach with spin-polarized scanning-tunneling spectroscopy only. \cite{Meir-Wiesendanger, Wiebe-Wiesendanger, Zhou-Wiesendanger} 

\begin{figure}[hbpt]
\centerline{\includegraphics[width=0.4\textwidth]{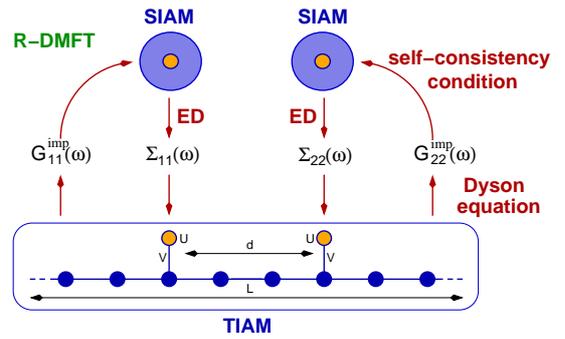}  }
\caption{(Color online) Schematic picture of real-space dynamical mean-field theory (R-DMFT) for the two-impurity Anderson model (TIAM). 
The system is given by two ``magnetic'' sites with strong Hubbard interaction $U$ (orange) at a distance $d$ coupled via a hybridization term of strength $V$ to a one-dimensional ``substrate'' consisting of $L$ non-interacting sites (blue) with nearest-neighbor hopping $t=1$.
In the R-DMFT, the TIAM is self-consistently mapped onto two single-impurity Anderson models which are solved independently by means of exact diagonalization (ED) to get the local self-energies. 
These are used to set up the TIAM Dyson equation the solution of which gives the local Green's functions which define via the R-DMFT self-consistency conditions the parameters of the impurity models
(see text for details).
}
\label{fig:Schematic_TIAM}
\end{figure}

The most simple model which captures this competition is displayed schematically in Fig.~\ref{fig:Schematic_TIAM}.
Here the electronic and magnetic properties of a magnetic adatom are modeled by a single non-degenerate orbital. 
A local magnetic moment is formed by a strong local Hubbard interaction. 
The adatom orbital hybridizes with a valence orbital of the nearest-neighboring substrate atom.
The substrate electronic structure itself is modeled by a tight-binding valence band resulting from non-degenerate and uncorrelated orbitals on a bipartite lattice with nearest-neighbor hopping.
Considering two adatoms yields a variant of the two-impurity Anderson model (TIAM) \cite{Alexander-Anderson_TIAM_first} in a surface geometry.
The main goal of our study is to benchmark the real-space variant \cite{RDMFT-Potthoff} of the dynamical mean-field theory \cite{Metzner,DMFT} that can be employed for theoretical studies of the electronic and magnetic properties of a single, of two and or of more magnetic adatoms in different geometries on metal surfaces.
For this purpose the TIAM represents a fundamental starting point.
In the Kondo limit of the TIAM, charge fluctuations on the adatom site are largely suppressed, and the adatom spin $\ff S^{f}$ couples antiferromagnetically to the local spin at the nearest-neighboring substrate site $\ff S^{c}$ via a spin-spin coupling $-J\ff S^{f} \ff S^{c}$ given by the local exchange $J\propto - V^{2}/U<0$.

The interplay between the Kondo effect and the RKKY interaction has extensively been studied in the Kondo limit of the TIAM 
or in the two-impurity Kondo model by different analytical as well as numerical techniques.
\cite{Alexander-Anderson_TIAM_first,Doniach,Perturbative_scaling,QMC_Fye87,NRG_Jones88,QMC_Fye89,NRG_Jones89,NRG_Sakai90,NRG_Sakai92,NRG_Ingersent,Exact_critical_Affleck92,Perturbative_U_Santro,QMC_Fye94,Exact_critical_Affleck95,Bosonization,DMRG_Hallberg,NRG_Vojta,DMRG_Nishimoto,OKWK07,QMC_Hoshino,LCL+10,Lichtenstein,JGS11}
The physics is governed by two energy scales, the nonlocal indirect magnetic interaction $J_{\rm RKKY}\propto J^{2}$ and the Kondo temperature $T_{K} \propto \exp(-1/|J|)$ below which the magnetic moment of the adatom (impurity) is screened locally.
In the Kondo regime for $T_{K} \gg |J_{\rm RKKY}|$, the conventional picture is that the local magnetic moments at the two impurities are individually screened by forming local singlet states with two Kondo clouds of itinerant electron spins from the substrate (conduction band).
For large $|J_{\rm RKKY}|$, on the other hand, and in the antiferromagnetic case $J_{\rm RKKY}<0$, the two adatom spins form a nonlocal singlet state and there is no Kondo effect.
If $J_{\rm RKKY}>0$ is ferromagnetic and large as compared to $T_{K}$, a nonlocal spin-triplet state is formed. 
This may subsequently be Kondo screened.
In the generic case and as a function of $J$ there is no quantum phase transition but a smooth crossover from the RKKY regime at weak $J$ to the Kondo regime at strong $J$.
For a dense system, i.e.\ the Kondo or Anderson lattice model, a static mean-field approach would sharpen this to a phase transition. \cite{Doniach}

Dynamical mean-field theory is a comprehensive, thermodynamically consistent and non-perturbative approximation for correlated lattice-fermion models. \cite{DMFT}
DMFT treats the Kondo effect exactly.
On the other hand, one has to tolerate an approximate treatment of the effects of the RKKY interaction.
It is important to note that there is no approximation of RKKY coupling itself:
Integrating out the non-interacting substrate degrees of freedom, the effective second-order-in-$J$ RKKY coupling, 
$J_{\rm RKKY, ij} = J^{2} \chi^{0,\rm sub}_{ij}(\omega=0)$, is given in terms of the nonlocal static susceptibility of the substrate.
It was pointed out by Peters and Pruschke \cite{PP07} that $J_{\rm RKKY, ij}$ is still finite but reduces to an interaction between nearest neighbors for the case of a lattice in infinite spatial dimensions where the DMFT becomes exact.
For finite dimensions, it is a long-ranged and oscillating function of the distance $d=|i-j|$.
What is neglected in fact for a finite-dimensional lattice, is the feedback of nonlocal, e.g.\ magnetic correlations, which result from the nonlocal RKKY coupling, on the self-energy and thus on the one-particle Green's function.
This is a rather subtle approximation the quality of which can be estimated by concrete numerical calculations only.

The same argumentation holds for the real-space DMFT (R-DMFT). \cite{RDMFT-Potthoff}
R-DMFT generalizes the standard DMFT to systems with missing or reduced translational symmetry by self-consistently mapping the original (lattice) model to a set of single-impurity Anderson models (SIAM), one for each of the geometrically or electronically inequivalent sites.
Even for the TIAM (see Fig.\ \ref{fig:Schematic_TIAM}), this real-space generalization is necessary if one wants to apply DMFT in order to test the local approximation for the self-energy.

Previous applications of the R-DMFT concentrated on the Mott metal-insulator transition at surfaces and in thin films,\cite{RDMFT-Potthoff,RDMFT-Mott} 
on surface effects in correlated Fermi liquids,
\cite{RDMFT-FL}
on multilayered nanostructures, heterostructures and interfaces,
\cite{RDMFT-nano} 
on disordered systems,
\cite{RDMFT-disorder} 
as well as on ultracold atomic gases in optical lattices with harmonic confinement.
\cite{RDMFT-ultracold}
It has not been employed, however, to study the effects of the indirect magnetic exchange.

The main purpose of the present study is to apply the R-DMFT to the particle-hole symmetric TIAM at half-filling and zero temperature and to study the magnetic response, i.e.\ different static magnetic susceptibilities, by applying a weak local magnetic field to one of the adatoms. 
Calculations are performed as a function of the distance between the adatoms and as a function of the hybridization strength $V$ to cross over from the Kondo to the RKKY regime.
To test the reliability of the dynamical mean-field approach, the substrate electronic structure is modeled as a one-dimensional tight-binding chain (see Fig.~\ref{fig:Schematic_TIAM}).  
The resulting essentially one-dimensional model is accessible to the density-matrix renormalization group (DMRG). 
\cite{Whi92,DMRG_Schollwoeck,DMRG_Verstraete}
Extensive comparison with numerically exact DMRG results obtained from an implementation based on matrix-product states, \cite{OR95} and operators along the lines described in Ref.\ \onlinecite{DMRG_McCulloch} helps to benchmark the mean-field approach. 

Our intention is that, by comparing with DMRG, the strengths but also the mean-field artifacts of R-DMFT become more transparent. 
A failure of R-DMFT for the weak-coupling limit, where non-local correlations due to the RKKY coupling are strong, can be expected from the very beginning. 
However, there are several interesting questions left, e.g.:
Where precisely are the limits of the mean-field approach?
How does a failure of the approach manifest itself in the observables?
Which physical effects are accessible to a description on the R-DMFT level?
To what extent can the physics be reproduced quantitatively in the strong-coupling limit?
Such benchmarking of the R-DMFT, at the level of the two-impurity Anderson model, will be important for future studies of similar systems in higher spatial dimensions, with more correlated adatoms forming more complex geometries such as chains or clusters etc.
By choosing the one-dimensional two-impurity Anderson model at half-filling, the above-mentioned questions are tackled in a situation that is very unfavorable for R-DMFT.
The benchmark will thus serve as a ``lower limit'' for the applicability of R-DMFT for future applications.

Our interest in the R-DMFT approach to study magnetic nanostructures on surfaces results from its extremely large flexibility. 
Opposed to DMRG, for example, the R-DMFT is able to investigate inhomogeneous systems in arbitrary geometries in higher dimensions. 
While this is actually characteristic for any mean-field approach, the R-DMFT is distinguished by the fact that it is non-perturbative and thermodynamically consistent.
To account for the effects of short-range correlations, the theory can be improved by certain cluster extensions, such as cellular DMFT. \cite{LCK08} 
This is, in principle, also conceivable for complicated inhomogeneous geometries but requires further methodical advances as there is no straightforward tiling of the lattice in most cases.

The paper is organized as the follows: 
The next sections introduces the model, notations and quantities of interest.
Sec.\ \ref{rdmft} and Sec.\ \ref{dmrg} briefly describe our real-space DMFT and our DMRG approach to the problem, respectively. 
Results of both approaches are presented, compared and discussed in detail in Sec.\ \ref{results}.
Finally, Sec.\ \ref{con} concludes the paper.

\section{MODEL AND BASIC THEORY}
\label{theory}

The Hamiltonian of the two-impurity Anderson model \cite{Alexander-Anderson_TIAM_first} displayed in Fig.\ \ref{fig:Schematic_TIAM} is given by:
\begin{eqnarray}
{\cal H}&=&-t\sum_{\langle i,j \rangle,\sigma}c_{i,\sigma}^\dagger c_{j,\sigma}^{\phantom\dagger}+U\sum_{\alpha=1}^{2}n_{\alpha,\uparrow}^f n_{\alpha,\downarrow}^f+\varepsilon\sum_{\alpha=1}^{2}n_{\alpha}^f \nonumber \\
&+&V \sum_{\alpha=1}^{2} \sum_\sigma\left(f_{\alpha,\sigma}^\dagger c_{i_{\alpha},\sigma}^{\phantom\dagger}+ h.c\right)-\mu\Bigl(\sum_{\alpha=1}^{2}n_{\alpha}^f+\sum_{i=1}^{L} n_{i}^c \Bigl) \: . \nonumber \\
\label{Hamiltonian}
\end{eqnarray}
Here  $f_{\alpha,\sigma}^\dagger$ and $c_{i,\sigma}^\dagger$ create an electron with spin projection $\sigma=\uparrow,\downarrow$ at the adatom sites $\alpha=1,2$ or at the substrate sites $i=1,2,\ldots,L$, respectively. 
$n_{\alpha,\sigma}^f= f_{\alpha,\sigma}^\dagger f_{\alpha,\sigma}^{\phantom\dagger}$ and $n_{i,\sigma}^c= c_{i,\sigma}^\dagger c_{i,\sigma}^{\phantom\dagger}$ denote the corresponding occupation-number operators.
The spin-summed occupation at one of the adatom sites and at one of the substrate sites are given by 
$n_{\alpha}^f=n_{\alpha,\uparrow}^f+n_{\alpha,\downarrow}^f$ and $n_{i}^c=n_{i,\uparrow}^c +n_{i,\downarrow}^c$, respectively.
The hopping amplitude $t$ between neighboring substrate lattice sites is used to fix the energy unit, i.e.\ $t=1$.
$V$ is the hybridization between an adatom site $\alpha$ and the nearest-neighbored substrate lattice site which is denoted by $i_\alpha$. 
$U$ and $\varepsilon$ are the on-site Hubbard interaction and the local on-site energy for the adatom sites. 
$\mu$ is chemical potential. 
In all our calculations we consider the particle-hole symmetric case with $\mu=0$ and $\varepsilon = -U/2$ where the system is half-filled, i.e.\ where the average occupation numbers in thermal equilibrium are given by
$\langle n_{\alpha}^f \rangle =1$ and $\langle n_{i}^c \rangle = 1$ for both $\alpha$ and all $i$.

The magnetic properties of the system are best characterized by site-dependent local and nonlocal susceptibilities. 
We consider the adatom-adatom susceptibilities,
\begin{eqnarray}
\label{imp_susceptibility}
\chi_{\alpha\beta}=\left.\frac{\partial m_\alpha^f}{\partial h_\beta}\right|_{h_\beta=0}=-\int_0^{1/T} d\tau \langle S_{\alpha,z}^f(\tau) S_{\beta,z}^f(0)\rangle \; ,
\end{eqnarray}
i.e.\ the local adatom susceptibilities $\chi_{\alpha\alpha}$ for $\alpha=1,2$ and the inter-adatom susceptibility $\chi_{12}=\chi_{21}$.
These provide information on the local adatom magnetic moment and, most important, on the indirect magnetic coupling.
Further, we are interested in the linear magnetic response of the substrate which is accessible via the adatom-substrate susceptibilities
 \begin{eqnarray}
\label{sub_susceptibility}
\chi_{i\beta}^{\rm sub}=\left.\frac{\partial m_i^c}{\partial h_\beta}\right|_{h_\beta=0}=-\int_0^{1/T} d\tau \langle S_{i,z}^c(\tau) S_{\beta,z}^f(0) \rangle \: .
\end{eqnarray}
Here $m_\alpha^f=\langle S_{\alpha,z}^f \rangle$ and  $m_i^c=\langle S_{i,z}^c \rangle$, with $S^f_{\alpha,z}=\frac{1}{2}(n_{\alpha,\uparrow}^f-n_{\alpha,\downarrow}^f)$ and 
$S_{i,z}^c=\frac{1}{2}(n_{i,\uparrow}^c-n_{i,\downarrow}^c)$, are magnetic moments on the adatom site $\alpha$ and on the substrate lattice site $i$ respectively. 
Furthermore, the imaginary-time dependence of an operator $A$ is given by $A(\tau) = e^{{\cal H} \tau} A e^{-{\cal H} \tau}$
In our calculations the susceptibilities, Eqs.\ (\ref{imp_susceptibility}) and (\ref{sub_susceptibility}), are computed as a numerical derivative with respect to a local magnetic field of strength $h_\beta$ coupling as ${\cal H}\rightarrow{\cal H}-h_\beta S_{\beta,z}$ to the Hamiltonian. 
Calculations are done for zero temperature in the present study.
Nevertheless, the formalism is set up for arbitrary finite $T$ below.

The main task is to compute, for a finite but weak field $h_{\beta}$, the spin-dependent average occupation numbers. 
These can be obtained via
\begin{eqnarray}
&&\langle n^f_{\alpha,\sigma} \rangle =\frac{1}{2}+2T\sum_{n\geq 0} \mbox{Re} \: G^{\rm imp}_{\alpha\alpha,\sigma}(i\omega_n)  \; , \\
&&\langle n^c_{i,\sigma} \rangle =\frac{1}{2}+2T\sum_{n\geq 0} \mbox{Re} \: G^{\rm sub}_{ii,\sigma}(i\omega_n)  
\end{eqnarray}
from the local single-electron adatom Green's function
$G^{\rm imp}_{\alpha\alpha,\sigma}(i\omega_n)=\langle\langle f_{\alpha,\sigma}^{\phantom\dagger};f_{\alpha,\sigma}^{\dagger}\rangle\rangle_{\omega_n}$ 
and the local substrate Green's function
$G^{\rm sub}_{ii,\sigma}(i\omega_n)=\langle\langle c_{i,\sigma}^{\phantom\dagger};c_{i,\sigma}^{\dagger}\rangle\rangle_{\omega_n}$ given at the fermionic Matsubara frequencies $\omega_n=(2n+1)\pi T$. 
The local Green's functions are the diagonal elements of the Green's function matrix $\hat G_\sigma(i\omega_n)$.
The latter can be obtained from the real-space Dyson equation:
\begin{equation}
\label{Green1}
\hat G^{-1}_\sigma(i\omega_n)=(i\omega_n+\mu) \hat I- \hat \varepsilon_{\sigma} - \hat {\cal T} - \hat \Sigma_\sigma(i\omega_n) \: ,
\end{equation}
where $\hat I$ is the unity matrix, $\hat \varepsilon_{\sigma}$ the diagonal local energy matrix, and $\hat {\cal T}$ is the hopping matrix.
$\hat \varepsilon_{\sigma}$ also includes the field term and is thus possibly spin-dependent.
$\hat {\cal T}$ not only includes the hopping $t$ between substrate sites but also hopping $V$ between the substrate and the adatom sites.
For a system with $L$ substrate sites and two adatoms, the matrix dimension is $L+2$ for each spin direction $\sigma$.

As there is a local Hubbard interaction on the adatom sites only, the self-energy $\hat \Sigma_\sigma(i\omega_n)$ is a $L+2$-dimensional matrix with non-zero elements $\Sigma_{\alpha\beta,\sigma}(i\omega_n)$ in the $2\times 2$ adatom-sites block only.
Hence, Eq.\ (\ref{Green1}) can be written as:
\begin{equation}
\label{Green2}
\hat G_\sigma(i\omega_n)=\left(
\begin{array}{c|c}
\begin{array}{cc}
\zeta_{1,\sigma} & -\Sigma_{12,\sigma} \\
-\Sigma_{21,\sigma} & \zeta_{2,\sigma}
\end{array} & \hat V \\
\hline
\hat V^\dagger & ({\hat{G}^0}){}^{-1}
\end{array}
\right)^{-1} \: ,
\end{equation}
where $\zeta_{\alpha,\sigma}=\zeta_{\alpha,\sigma}(i\omega_n)=i\omega_n+\mu-\varepsilon_{\sigma} -\Sigma_{\alpha\alpha,\sigma}(i\omega_n)$, and where
$\hat V$ is the $2\times L$ hybridization matrix including hopping between adatoms and substrate only. 
Its non-zero elements are given by $V_{1,i_1}=V_{2,i_2}=V$. 
Further, $\hat G^0$ is the non-interacting substrate Green's function matrix. 
In case of periodic boundary conditions, its elements are
\begin{eqnarray}
G^0_{i j}(i\omega_n)=\frac{1}{L}\sum_{m=0}^{L-1}\frac{\cos\left(k_m (i-j)\right)}{i\omega_n+\mu-\varepsilon(k_m)}
\end{eqnarray}
where $k_m=2\pi m/L$ with $m =0,1,\ldots,L-1$, while for open boundary conditions
\begin{eqnarray}
G^0_{i j}(i\omega_n)=\frac{2}{L+1}\sum_{m=1}^{L}\frac{\sin(k_m i) \sin(k_m j)}{i\omega_n+\mu-\varepsilon(k_m)} 
\end{eqnarray}
where $k_m=\pi m/(L+1)$ with $m=1,2,\ldots,L$. 
In both cases $\varepsilon(k)=-2t\cos(k)$ is non-interacting dispersion.

Generally, to calculate, for a given self-energy, the local Green's functions, one has to numerically invert the matrix given by Eq.\ (\ref{Green1}).
In our case for the TIAM, however, it is possible to ``integrate out'' the substrate degrees of freedom and to find analytical expressions which substantially reduce the numerical effort:
Using the identity 
\begin{eqnarray}
\label{matrix_inv} 
\left(
\begin{array}{cc}
\hat A & \hat B \\
\hat C & \hat D
\end{array}
\right)^{-1}\hspace{-0.4cm}=
\left(
\begin{array}{cc}
\hat F &   -\hat F \hat B \hat D^{-1}\\
-\hat D^{-1} \hat C \hat F &  \hspace{0.1cm}\hat D^{-1}\hspace{-0.1cm} +\hspace{-0.1cm} \hat D^{-1} \hat C \hat F \hat B \hat D^{-1}
\end{array}
\right)
\end{eqnarray}
with $\hat F=\left(\hat A- \hat B \hat D^{-1} \hat C\right)^{-1}$ which is valid for arbitrary quadratic matrices $\hat A$ and $\hat D$ and arbitrary rectangular matrices $\hat C$ and $\hat D$,
we find
\begin{equation}
\label{Green3}
\hat G_\sigma(i\omega_n)=
\left(
\begin{array}{cc}
\hat G^{\rm imp}_\sigma &  
- \hat G^{{\rm imp}}_\sigma \hat V \hat G^0 \\
-\hat G^0 \hat V^\dagger \hat G^{{\rm imp}}_\sigma & \hat G^{\rm sub}_\sigma
\end{array}
\right) \; ,
\end{equation}
where 
\begin{eqnarray}
\label{Green_Gimp}
\hat G^{{\rm imp}}_\sigma(i\omega_n)=
\left(\left(
\begin{array}{cc}
\zeta_{1,\sigma} & -\Sigma_{12,\sigma} \\
-\Sigma_{21,\sigma} & \zeta_{2,\sigma}
\end{array}
\right)-\hat V \hat G^0 \hat V^\dagger
\right)^{-1} 
\end{eqnarray}
is the $2 \times 2$ adatom Green's function matrix and
\begin{eqnarray}
\label{Green_Gsub}
\hat G^{\rm sub}_\sigma=\hat G^0 + \hat G^0 \hat V^\dagger \hat G^{{\rm imp}}_\sigma \hat V \hat G^0 
\end{eqnarray}
is the $L \times L$ substrate Green's function matrix.
The remaining task thus consists in the inversion of a $2\times 2$ matrix 
\begin{equation}
\label{Green_matrix}
\hat G^{{\rm imp}}_\sigma(i\omega_n)=\left(
\begin{array}{cc}
\zeta_{1,\sigma}-\Delta_{11}  & -\Sigma_{12,\sigma}-\Delta_{12}  \\
-\Sigma_{21}-\Delta_{21,\sigma}  & \zeta_{2,\sigma}(i\omega_n)-\Delta_{22}
\end{array}
\right)^{-1} \; ,
\end{equation}
where 
\begin{equation}
\Delta_{\alpha\beta}(i\omega_n)=V G^0_{i_\alpha i_\beta}(i\omega_n) V
\label{hyb}
\end{equation}
is the hybridization function.
This is readily done:
\widetext
\begin{eqnarray}
\label{gimp1}
&&G_{11,\sigma}^{\rm imp}(i\omega_n)=
\frac{\zeta_{2,\sigma}(i\omega_n)-\Delta_{22}(i\omega_n)}
{(\zeta_{1,\sigma}(i\omega_n)-\Delta_{11}(i\omega_n))(\zeta_{2,\sigma}(i\omega_n)-\Delta_{22}(i\omega_n))-
(\Delta_{12}(i\omega_n)+\Sigma_{12,\sigma}(i\omega_n))^2} \; , \\
\label{gimp2}
&&G_{22,\sigma}^{\rm imp}(i\omega_n)= \frac{\zeta_{1,\sigma}(i\omega_n)-\Delta_{11}(i\omega_n)}
{(\zeta_{1,\sigma}(i\omega_n)-\Delta_{11}(i\omega_n))(\zeta_{2,\sigma}(i\omega_n)-\Delta_{22}(i\omega_n))-
(\Delta_{12}(i\omega_n)+\Sigma_{12,\sigma}(i\omega_n))^2}\; ,\\
\label{gimp3}
&&G_{12,\sigma}^{\rm imp}(i\omega_n)= G_{21,\sigma}^{\rm imp}(i\omega_n)=\frac{\Delta_{12}(i\omega_n)+\Sigma_{12,\sigma}(i\omega_n)}
{(\zeta_{1,\sigma}(i\omega_n)-\Delta_{11}(i\omega_n))(\zeta_{2,\sigma}(i\omega_n)-\Delta_{22}(i\omega_n))-
(\Delta_{12}(i\omega_n)+\Sigma_{12,\sigma}(i\omega_n))^2}\; .
\end{eqnarray}
This provides us with the local adatom Green's functions in particular and, using Eq.\ (\ref{Green_Gsub}), with the local Green's functions for each substrate site via:
\begin{equation}
\label{Gsub}
G^{\rm sub}_{ii,\sigma}(i\omega_n)
=
G^0_{ii}(i\omega_n)\\
+ 
\sum_{\alpha,\beta} G^0_{ii_\alpha}(i\omega_n) V G_{\alpha\beta,\sigma}^{\rm imp}(i\omega_n) V G^0_{i_\beta i} (i\omega_n) \; .
\end{equation}
\endwidetext

\section{Real-space dynamical mean-field theory}
\label{rdmft}

To complete the theory, we need the self-energy matrix $\Sigma_{\alpha\beta,\sigma}(i\omega_n)$. 
This requires an approximation. 
Within real-space DMFT, \cite{RDMFT-Potthoff} the self-energy is obtained by considering weak-coupling perturbation theory in $U$ to all orders and by summing all local diagrams in the skeleton-diagram expansion of the self-energy, $\hat{\Sigma} = \hat{\Sigma}[\hat G]$.
This implies that the resulting self-energy is local: $\Sigma_{\alpha\beta,\sigma}(i\omega_n) = \delta_{\alpha \beta} \Sigma_{\alpha,\sigma}(i\omega_n)$ but possibly site-dependent.
For correlated lattice models with full translational symmetries, the approach reduces to the conventional DMFT. \cite{DMFT}
As in the conventional DMFT, the local diagrams are not summed explicitly, the problem is rather reformulated by introducing a self-consistent mapping onto an effective single-impurity problem.
Here, however, the self-consistent cycle is more complicated since a lattice model with $M$ geometrically or electronically inequivalent sites has to be self-consistently mapped onto a set of $M$ effective single-impurity models. 
In our case we have to consider at most $M=2$ single-impurity Anderson models (see Fig.\ \ref{fig:Schematic_TIAM}): 

We start with a guess for the local self-energies $\Sigma_{\alpha,\sigma}(i\omega_n)$, i.e.\ for $\zeta_{\alpha,\sigma}(i\omega_n)$.
This is used in the Dyson equation of the lattice model to compute the Green's function matrix, and in particular the local elements of the Green's function matrix at the correlated sites.
In our case, we can profit from Eqs.\ (\ref{gimp1}) and (\ref{gimp2}) to get the local adatom Green's functions $G_{\alpha\alpha,\sigma}^{\rm imp}(i\omega_n)$ directly.
The R-DMFT self-consistency conditions,
\begin{equation}
\label{localDyson}
\frac{1}{
{\cal G}^0_{\alpha,\sigma}(i\omega_n)} 
= 
\frac{1}{G_{\alpha\alpha,\sigma}^{\rm imp}(i\omega_n)}
+
\Sigma_{\alpha,\sigma}(i\omega_n) \; ,
\end{equation}
then provide us with the Weiss Green's functions ${\cal G}^0_{\alpha,\sigma}(i\omega_n)$ for $\alpha=1,...,M$, i.e.\ with the non-interacting Green's functions of the $M$ effective impurity models.
These can be written as 
${\cal G}^0_{\alpha,\sigma}(i\omega_n) = i \omega_n +\mu - \varepsilon - \Delta_\alpha(i\omega_n)$.
The one-particle parameters of each effective SIAM, the one-particle energies of the bath sites as well as the corresponding hybridization strengths, are found from the poles and the residues of the corresponding hybridization function $\Delta_\alpha(i\omega_n)$ (which should not be mixed up with $\Delta_{\alpha\beta}(i\omega_n)$, see Eq.\ (\ref{hyb})).
Once the effective impurity models are fixed, the crucial step consists in the solution of the models which can be done independently for any $\alpha=1,...,M$.
This yields the self-energies $\Sigma_{\alpha,\sigma}(i\omega_n)$ and thus closes the self-consistency cycle (Fig.\ \ref{fig:Schematic_TIAM}).
This procedure is iterated until converged self-energies are obtained.

As an impurity solver to get the self-energy $\Sigma_{\alpha,\sigma}(i\omega_n)$ of the $\alpha$-th SIAM we use the exact-diagonalization (ED) method. \cite{ED-Caffarel,ED-Si} 
Here, a finite small number $n_s-1$ of auxiliary bath degrees of freedom in the effective SIAM is considered.
We use full diagonalization with $n_s=6$ and the Lanczos method \cite{Lanczos-Golub,Lanczos-Koch} with $n_s=8$ and $n_s=10$.
Exploiting the fact that the total particle number and the $z$-component of the total spin are conserved quantities, the diagonalization can be done in smaller invariant subspaces of the full Hilbert space. 
All calculations have been done, if not stated differently, with $n_s=10$.
For a given Weiss Green's function, the one-particle bath parameters of the SIAM are found by a minimization procedure on the imaginary-frequency axis as described in Ref.\ \onlinecite{ED-Caffarel} using high-frequency cutoff of the order of $U$ and low-frequency cutoff specified by the {\em fictitious} temperature $T/t=0.001$.
With the latter we can formally work in the finite-temperature Matsubara framework as outlined above. 
On the other and, the value of the fictitious temperature chosen is clearly lower than the smallest energy scale that can be accessed by means of the ED solver for $n_{s}=10$. 
We have regularly checked that the results do not significantly depend on the cutoff and on $n_{s}$.

The computational effort of the R-DMFT scheme roughly scales linearly with the number of impurity models, i.e.\ with the number of inequivalent sites in the original system. 
While here we focus on the $M=2$ case for benchmarking purposes, future applications are intended that address systems with up to ${\cal O}(100)$ inequivalent magnetic atoms. 
We expect that those applications can still be performed conveniently using ED as a solver.
It might nevertheless be interesting to have a scheme at hand that is considerably faster.
Here, the two-site DMFT \cite{Pot01} represents an alternative. 
The main idea is the use a single bath degree of freedom only, as in the so-called linearized DMFT \cite{BP00} for the Mott-Hubbard transition. 
The resulting effective two-site impurity model is readily solved.
On the other hand, the DMFT self-consistency condition can obviously no longer be satisfied exactly or to a high level of accuracy as in the ED approach with, say, $n_s=10$ sites.
It has therefore been suggested \cite{Pot01} to take into account the leading orders in systematic expansions of the self-consistency condition for high and for low frequencies only.
This results in a simplified but extremely fast approach which is suitable to get a quick overview of magnetic phase diagrams, for example.
The real-space extension of two-site DMFT to systems with reduced translational symmetries is straightforward and will be discussed in detail elsewhere. 

\section{Density-matrix renormalization}
\label{dmrg}

The two-impurity Anderson model in an essentially one-dimensional geometry (see Fig.\ \ref{fig:Schematic_TIAM}) is amenable to a numerically exact solution by using the density-matrix renormalization group. \cite{Whi92,DMRG_Schollwoeck,DMRG_Verstraete}
Therefore, DMRG calculations can be used to benchmark the quality of the magnetic susceptibilities obtained from the R-DMFT approach.
The calculation of the ground state and of ground-state expectation values for the TIAM is a standard problem within DMRG. 
Here we have been following Ref.\ \onlinecite{DMRG_McCulloch} and have implemented a code which is based on the variation of matrix-product states (MPS). \cite{OR95} 
The main idea is to optimize a test wave function $|\psi\rangle$ of the form 
\begin{align}
   |\psi\rangle = \sum_{n_1,\ldots,n_L} \mathbf A^{(n_1)} \ldots \mathbf A^{(n_L)} |n_1\rangle \ldots |n_L\rangle \: ,
\end{align}
where $\{ |n_q \rangle \}$ is a local basis at the site $q$ of a one-dimensional chain with $L$ sites in total.
The elements $A_{i_{q-1}i_q}^{(n_q)}$ of the matrices $\mathbf A^{(n_q)}$ are considered as variational parameters which are locally and iteratively optimized during a sweep through the chain by exploiting the Ritz variational principle.
Several sweeps are necessary to obtain a converged ground state.

In practice, the local optimization can be reformulated as a generalized eigenvalue problem which is simplified to an ordinary one by exploiting a local gauge invariance of $|\psi\rangle$ to properly (left- and right-) orthogonalize the $A$-matrices. 
The eigenvalue problem is then efficiently solved by means of the Davidson method. \cite{Dav75}
We profit from the so-called wave-function transformation \cite{Whi96} to reduce the number of iterations necessary for convergence of the Davidson algorithm and exploit the two $U(1)$ symmetries of the Hamiltonian corresponding to conservation of the total particle number and the $z$-component of the total spin. 

Observables and the Hamiltonian in particular are represented as matrix-product operators. \cite{DMRG_McCulloch}
Besides an elegant and flexible coding this allows to easily consider different implementations of the Hamiltonian.
For the present case of the TIAM there are two possibilities to treat the adatoms suggesting themselves: 
(i) An adatom orbital $\alpha$ and the substrate orbital $i_\alpha$ ``below'' $\alpha$ are treated as a single ``site'' $q$ in the DMRG context.
The disadvantage is that therewith the local Hilbert-space dimension at $q$ is enlarged.
(ii) The adatom orbitals $\alpha$ are treated as separate sites, i.e.\ a chain of length $L+2$ is formed.
This leaves the local Hilbert-space dimension constant but introduces next-nearest-neighbor hopping terms.
We have tested both variants and found the differences in computational costs and accuracy to be marginal only.
Routinely, variant (ii) is employed.

An important aspect is to prevent the sweep algorithm from getting stuck in a local energy minimum. 
This can be circumvented by implementing a mixed single-site approach \cite{Whi05,DMRG_McCulloch} to introduce fluctuations in the reduced density matrix. 
The additional coupling to a larger set of states considerably improves the convergence properties when optimizing $|\psi\rangle$. 
It furthermore also allows to dynamically adapt the dimensions of the $A$-matrices.
Converged results for typical situations with long-range spin-spin correlations in a TIAM with about $L=50$ sites are obtained with matrix dimensions of the order of $m=400$ in the largest invariant blocks of the $A$-matrices.
A reliable error measure is the variance $r=\langle \psi | ({\cal H}-E)^2 | \psi \rangle$ which is easily accessible within an MPS-based implementation.
We have checked that typically the standard deviation $\sqrt r < 10^{-4}$.

\section{RESULTS}
\label{results}

\begin{figure*}[hbpt]
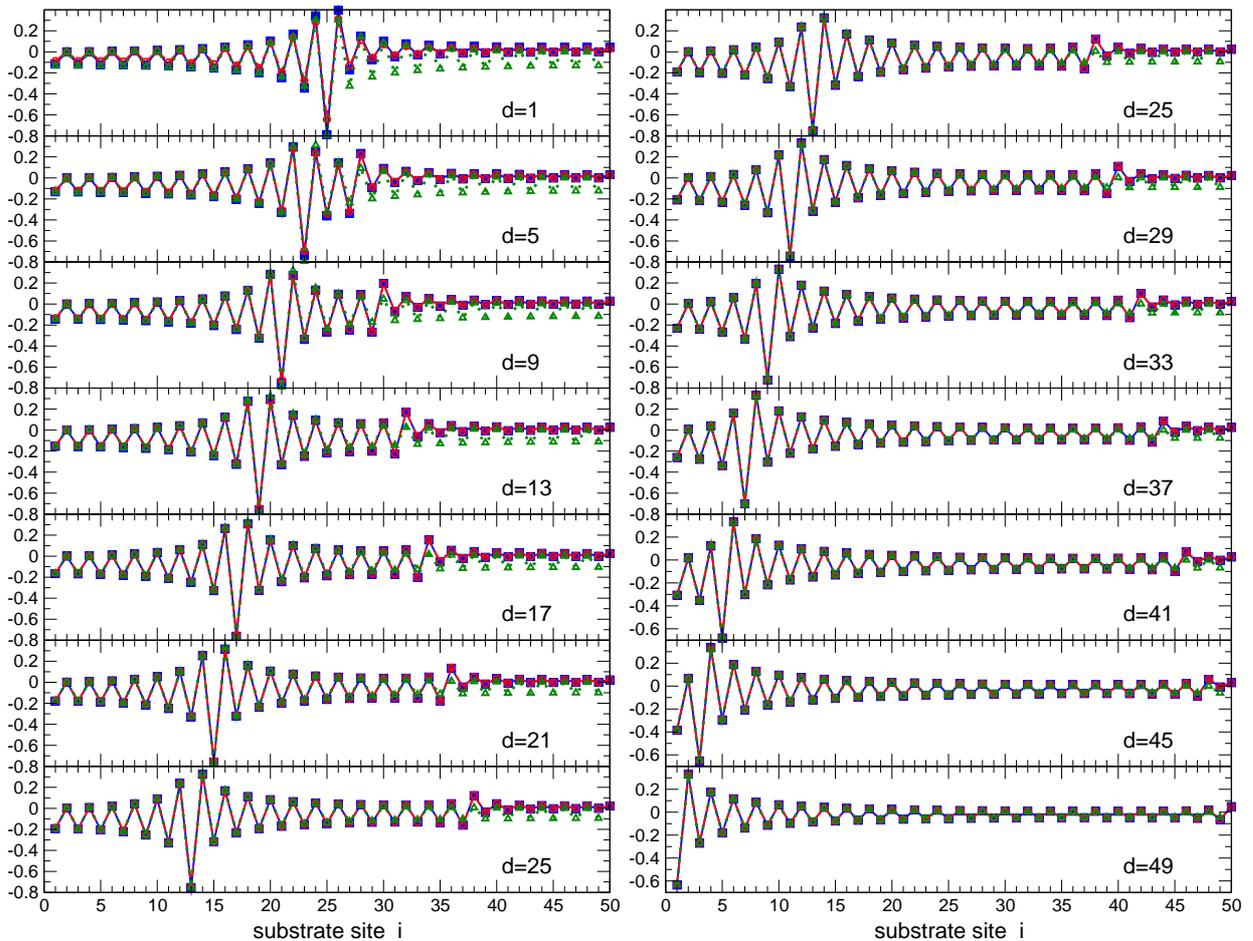

\includegraphics[width=0.45\textwidth]{DMFT_chi_1-25.eps}
\includegraphics[width=0.45\textwidth]{DMFT_chi_25-49.eps}
\caption{(Color online) Static magnetic susceptibility $\chi_{i1}^{\rm sub}$ at Hubbard interaction $U=8$ and hybridization strength $V=\sqrt{2}$ for a system with $L=50$ substrate sites and two adatoms at positions symmetric to the chain center and different distances $d$ as indicated. 
$\chi_{i1}^{\rm sub}$ gives the linear response of the substrate at site $i$ to a local magnetic field at the first (left) adatom.
Energy scale: nearest-neighbor hopping in the substrate $t=1$.
Blue lines with squares: results as obtained from real-space DMFT using exact diagonalization with $n_{s}=10$ as a solver. 
Red dashed lines with circles: numerically exact solution as obtained from DMRG calculations.
For comparison DMRG results for a system with $L=49$ sites and a single adatom are shown (dotted green line with triangles).
This corresponds to switching off the hybridization $V$ between the substrate and the second (right) adatom.
}
\label{Substrate_chi}
\end{figure*}

R-DMFT and DMRG calculations have been performed for the TIAM at half-filling and zero temperature.
We consider systems with an even number $L$ of substrate sites and two adatoms at positions symmetric to the chain center at a distance $d=|i_1-i_2|$ as displayed in Fig.\ \ref{fig:Schematic_TIAM}.
As this implies an even number of electrons, there is no Kramers degeneracy of the ground state.
All calculations are done using open boundary conditions. 

Fig.\ \ref{Substrate_chi} shows the magnetic susceptibility $\chi_{i\beta}^{\rm sub}$ for a TIAM with $L=50$ as defined in Eq.\ (\ref{sub_susceptibility}). 
Because of the mirror symmetry, it is sufficient to discuss e.g.\ the case $\beta=1$, i.e.\ the left adatom.
For the calculations we apply a weak local magnetic field with a strength $h_{\beta=1} = 10^{-5}$--$10^{-2}$ at the left adatom and look for the response at substrate site $i$.
Regularly, calculations for different $h_1$ are performed to ensure that the field strength is in the linear-response regime.

Let us first concentrate on distances $d=4n+1$ with integer $n$.
Here the RKKY coupling between the magnetic adatoms is antiferromagnetic.
Other distances $d$ including those with ferromagnetic coupling will be discussed in Sec.\ \ref{sec:diffdist}.

\subsection{Magnetic response of the substrate}

We start the discussion with $d=49$ (lowest panel on the right in Fig.\ \ref{Substrate_chi}). 
This is the case where the two adatoms are located at the edges of the substrate chain. 
The blue lines refer to our R-DMFT calculations which have been done with $n_s=10$ local degrees of freedom in the effective impurity model.
Directly ``below'' the first adatom at $i_1$ the response is antiferromagnetic, i.e.\ $\chi_{i_{1},1}<0$.
This simply reflects the antiferromagnetic Kondo coupling $J$.
The calculations have been done for $U=8$ and $V^2=2$ where the nearest-neighbor hopping in the substrate $t=1$ is used to set the energy scale.
This results in a negative, i.e.\ antiferromagnetic, local exchange interaction of intermediate strength $J=-8V^2/U = -2$.
This is clearly beyond the weak-coupling limit $J\to 0$ but still charge fluctuations are largely suppressed: 
We find an average double occupancy of $\langle n_{1\uparrow} n_{1\downarrow} \rangle = 0.072$ at the adatom site, and the adatom local magnetic moment $\langle \ff S_1^2 \rangle = 3 ( 1 - 2 \langle n_{1\uparrow} n_{1\downarrow} \rangle) / 4 = 0.64$ is much closer to the localized-spin value $3/4$ than to the free fermion value $3/8$.

As a function of the distance $|i-i_{1}|$ to the first impurity, the response is oscillatory corresponding to the $2k_F = \pi$ nesting wave vector.
Its absolute value is maximal at $i_1$, decreases with increasing $i$ and almost saturates until there is a slight upturn for $i\to i_2=50$, i.e.\ at the position of the second adatom. 
Consistent with the $2k_F$ oscillation, $\chi_{i,1}^{\rm sub}$ is positive at $i=i_2$ which implies, due to the antiferromagnetic local coupling $J<0$, that there is an antiferromagnetic (RKKY) alignment of the two adatom moments.

The corresponding DMRG results are also shown in Fig.\ \ref{Substrate_chi} for comparison (red lines).
For the distance $d=49$, however, there is actually no difference to the R-DMFT results visible on the scale of the figure. 
As R-DMFT accounts for the single-impurity Kondo effect exactly, this perfect agreement would be plausible if a picture of two independent Kondo effects applied. 
Strictly speaking, however, this cannot be the case: 
There is a finite nonlocal adatom-adatom susceptibility, even in this long-distance limit (see also Fig.\ \ref{DMFT_vs_DMRG_d} and corresponding discussion below) which in principle has a non-vanishing feedback on the self-energy and generates nonlocal elements of the self-energy in particular.
R-DMFT is thus approximate.
On the other hand, we can conclude that this feedback of the nonlocal susceptibility is apparently negligibly small and R-DMFT almost exact for the present situation.

Upon decreasing the distance between the adatoms, this picture should change gradually. 
However, apart from the extreme case $d=1$, deviations of the R-DMFT from the DMRG results are extremely small, and the agreement between R-DMFT and DMRG remains excellent.
On the other hand, with decreasing $d$, the $i$ dependence of the susceptibility becomes much more complicated:
The response below the second adatom (see the second maximum of $|\chi_{i,1}^{\rm sub}|$) becomes stronger and stronger, the response at substrate sites between the adatoms increases and its absolute value develops a pronounced minimum close to $i_2$, whereas the response beyond the second adatom, for $i>i_2$, gets very weak. 
Furthermore, while the susceptibility changes sign between nearest neighbors, its two-site {\em average} is negative between the adatoms and also beyond the first one for $i<i_1$ but is found to be positive for $i>i_2$.
There is another subtle observation, namely the (ferromagnetic) response at the nearest neighbor to the right of $i_1$ is larger than the one to the left of $i_1$ for all $d$ down to $d=1$, except for $d=5$ and $d=9$. 
The ratio $\chi_{i_1+1,1}^{\rm sub} / \chi_{i_1-1,1}^{\rm sub}$ is decreasing with decreasing $d$ becomes smaller than unity for $d=5$ and $d=9$ and larger than unity again for $d=1$.

All these non-trivial features are perfectly captured by the R-DMFT and in fact result from an effective adatom-adatom interaction. 
This becomes obvious by comparing the results for the TIAM with those of a corresponding single-impurity Anderson model where the second (right) adatom $\alpha=2$ is missing or, equivalently where the hybridization to the second adatom is switched off.
We have performed DMRG calculations for corresponding single-adatom models. 
To ensure a singlet ground state at half-filling, however, the substrate chain has to be shortened by one site on the right edge ($L=49$).
The resulting substrate susceptibilities $\chi_{i,1}^{\rm sub}$ are shown in Fig.\ \ref{Substrate_chi} as green lines.

Comparing the SIAM and the TIAM results to each other once more demonstrates that the effects of the indirect nonlocal RKKY coupling become more and more pronounced with decreasing $d$.
The differences between the single-adatom and the two-adatom physics visible in the susceptibilities for $i>i_2$ are larger by more than an order of magnitude than the differences between the R-DMFT and the DMRG results. 
Again this shows that there are sizable effects on nonlocal magnetic correlations which do not fully feed back to the one-electron self-energy.

\subsection{Spin correlations and nonlocal susceptibitlities}

\begin{figure}[t]
\includegraphics[width=0.45\textwidth]{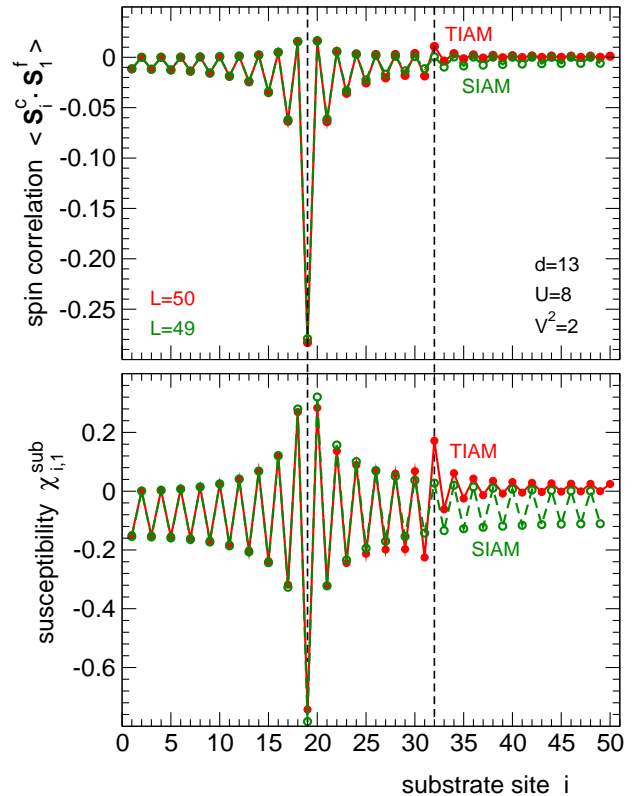}  
\caption{(Color online) Spin-spin correlation function $\langle {\bf S}_i^c {\bf S}_1^f \rangle$ (upper panel) and magnetic susceptibility $\chi_{i,1}^{\rm sub}$ (lower panel) for $U=8$ and $V=\sqrt{2}$ as obtained by DMRG for a system with $L=50$ ($L=49$) substrate sites and two adatoms (one adatom) as functions of the substrate site $i$.
Red lines: results for two adatoms and $L=50$, TIAM.
Green lines: results for $L=49$ and a single adatom at the same position as the $\beta=1$ (left) adatom in the two-adatom model, SIAM.
The dashed lines indicate the positions $i_{1}$ and $i_{2}$ of the substrate sites ``below'' the adatoms.
}
\label{Correlation_function}
\end{figure}

The DMRG data for $\chi_{i,1}^{\rm sub}$ and also for the equal-time spin-spin correlation function $\langle \ff S_1^f \ff S_i^c\rangle$ are shown in Fig.\ \ref{Correlation_function} for $d=13$ on a larger scale.
Let us discuss the physics of this situation in detail. 
The response of the substrate to a static local field at $\beta=1$ is governed by the low-energy excitations around the Fermi edge, i.e.\ $\omega=0$.
Contrary, the equal-time spin-spin correlation is obtained by a frequency integration of the dynamic (retarded) susceptibility $\chi_{i,1}^{\rm sub}(\omega)$ and thus includes several energy scales. 
Nevertheless, the spin-spin correlation behaves qualitatively very similar to $\chi_{i,1}^{\rm sub}$, and we will refer to this on an equal footing with the susceptibility.

The ground state of the whole system is a spin singlet in all calculations discussed here.
This provides us with a simple sum rule for the spin-spin correlation: 
Exploiting rotational symmetry, we have $\langle \ff S^{\rm tot} \ff S^{f}_1 \rangle = 3 \langle S^{\rm tot}_{z} S^{f}_{1z} \rangle$ where $\ff S^{\rm tot} = \ff S^{f}_{1} + \ff S^{f}_{2} + \ff S^{\rm sub}$ is the total spin and $\ff S^{\rm sub} = \sum_{i=1}^{L} \ff S^{c}_{i}$ the total substrate spin.
Using $M^{tot}=0$ in the ground state, we immediately find:
\begin{equation}
\langle \ff S^{f}_1 \ff S^{f}_1 \rangle + \langle \ff S^{f}_2 \ff S^{f}_1 \rangle + \langle \ff S^{\rm sub} \ff S^{f}_1 \rangle = 0 \: .
\label{sumrule}
\end{equation}
For the single-adatom model we then have 
\begin{equation}
  \langle \ff S^{f}_1 \ff S^{f}_1 \rangle  + \langle \ff S^{\rm sub} \ff S^{f}_1 \rangle = 0 \: .
\label{sumrulesiam}
\end{equation}
which explains why the spin-spin correlation is mainly negative: Namely, if summed over all substrate sites, it just compensates the adatom local moment.
For the two-adatom model at $d=13$, the adatom-adatom spin correlation $\langle \ff S^{f}_1 \ff S^{f}_2 \rangle$ is negative but its absolute value is small compared to $\langle \ff S^{f}_1 \ff S^{f}_1 \rangle$. 
Looking at Eq.\ (\ref{sumrule}), the overall substrate response is thus still antiferromagnetic but somewhat reduced as compared to the single-adatom model.

Qualitatively the same applies to the susceptibility as can be seen from the lower panel in Fig.\ \ref{Correlation_function}.
For any large but finite system with a non-degenerate singlet ground state, we again have a simple sum rule:
A singlet ground state and a finite gap implies that the total magnetic moment must vanish for any $h_{1}$ up to some finite critical field:  $\langle \ff S^{\rm tot} \rangle = \langle \ff S^{f}_{1} \rangle + \langle \ff S^{f}_{2} \rangle + \langle \ff S^{\rm sub} \rangle = 0$.
Taking the derivative with respect to $h_{1}$ then yields:
\begin{equation}
\chi_{11} + \chi_{21} + \sum_i \chi_{i1}^{\rm sub} = 0 \: .
\label{sumrulechi}
\end{equation}
In the same way as above, $\sum_i \chi_{i1}^{\rm sub} = - \chi_{11} < 0$ for a single adatom, and for two adatoms the total response of the substrate is still negative but slightly reduced in absolute magnitude due to the presence of the second adatom since $\chi_{21}<0$ at $d=13$.

As can be seen in Fig.\ \ref{Correlation_function} by comparing with the results for the single-adatom model, the most pronounced effect due the presence of the second adatom consists in the strong enhancement of $\chi_{i_2 1}$, i.e.\ the response below the second adatom.
This can easily be understood by referring to the RKKY limit for a system of finite size $L$:
For $V \to 0$ keeping $U\gg t$ fixed, charge fluctuations vanish and we are left with a Kondo-type model. 
In the weak-coupling limit $J \to 0$ the substrate degrees of freedom can be integrated out, and the adatom magnetic response, i.e.\ $\chi_{11}$ and $\chi_{21}$, is perfectly described by an effective RKKY two-spin model
\begin{equation}
  H_{\rm RKKY} = - J_{\rm RKKY} \ff S_1^f \ff S_2^f \: ,
\label{eq:rkky}
\end{equation}
where $J_{\rm RKKY} = J^2 \chi_{i_{1}i_{2}}^{0,\rm sub}$ is given in terms of the static substrate susceptibility at $J=0$.
The Kondo effect, on the other hand, does not interfere with this picture as it is cut by the finite-size gap: 
One can define a coupling strength $J_c$ at which the Kondo temperature $T_K$ becomes comparable with the finite-size gap. 
Then, for $J<J_c$ the Kondo effect is absent as there are simply no states at the Fermi energy available to screen the adatom moment. \cite{TKvD99}
This implies that the substrate is in a singlet state for weak $J$ and thus
$\chi_{1}^{\rm sub} \equiv \sum_i \chi_{i1}^{\rm sub} = (\partial / \partial h_{1}) \langle S_{z}^{\rm sub} \rangle = 0$, i.e.\ there is no substrate contribution to the magnetic moment induced by the field at $\beta=1$.
From the sum rule Eq.\ (\ref{sumrulechi}) we thus have $\chi_{11}+\chi_{21}=0$, i.e.\ also the two adatom spins form a perfect singlet consistent with Eq.\ (\ref{eq:rkky}).
Hence, applying a field $h_{1}$ at adatom $\beta=1$ induces antiferromagnetically aligned magnetic adatom moments with {\em the same} absolute magnitude.
For $J$ beyond but close to the RKKY limit we therefore expect the absolute magnitude of the substrate response at $i_{1}$ and $i_{2}$ as almost equal.
For finite and actually intermediate $J$, see Fig.\ \ref{Correlation_function}, the effect is strongly diminished but still clearly visible.
Note that the above argumentation can analogously be given by referring to the spin-spin correlation.

As mentioned before, looking at the sum rules (\ref{sumrule}) and (\ref{sumrulechi}), we can understand that the response of the substrate is somewhat attenuated in the TIAM as compared to the SIAM.
This reduction, however, is not homogeneous: 
There is a comparatively strong reduction beyond the second adatom for $i>i_2$ while the response is nearly the same or even somewhat enhanced close to $i_{2}$ for $i<i_2$, and there is almost no effect for $i<i_1$.
That the effect is least pronounced close to the first adatom, can easily be understood by referring to the extreme Kondo limit where a picture of two separate Kondo clouds applies.
In this case the magnetic response to the field applied to the first adatom would be the same as the response in the corresponding single-adatom model. 
As is seen in Fig.\ \ref{Correlation_function}, however, close to $i_1$ there are finite differences, i.e.\ the Kondo clouds do overlap, but the differences are small. 
Since according to the sum rule the total response must be weaker in the TIAM, a reduced response must and in fact does show up away from $i_1$, i.e.\ for $i>i_2$.

\begin{figure}[t]
\includegraphics[width=0.48\textwidth]{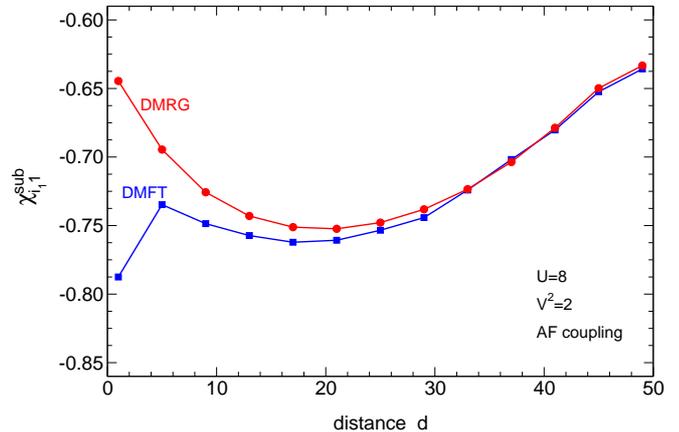}  
\caption{(Color online) Susceptibility $\chi_{i_1 1}^{\rm sub}$ at the lattice site $i_1$ below the first adatom as a function of the adatom-adatom distance $d$ for $U=8$, $V=\sqrt{2}$ and $L=50$ substrate sites as obtained from R-DMFT (blue line) and DMRG (red line). 
}
\label{local_substrate_chi}
\end{figure}

The sum rule (\ref{sumrulesiam}) for the SIAM may also be used to roughly estimate the size of the individual ``Kondo clouds''. 
Using the DMRG data for $\langle {\bf S}_1^f {\bf S}_i^c\rangle$, we define an integrated spin-spin correlation function, \cite{HCSD09}
\begin{equation}
\varTheta(r)=1+\sum_{|i-i_1| < r} \frac{\langle {\bf S}_1^f  {\bf S}_i^c \rangle}{\langle {\bf S}_1^f  {\bf S}_1^f \rangle} \: , 
\label{cloud}
\end{equation}
for the single-adatom model.
We have $\varTheta(0)=1$. 
With increasing $r$ more and more substrate spins around $i_{1}$ are included in the sum, and $\varTheta(r)$ essentially decreases with $r$ until $\varTheta=0$ if all spins are included as is obvious from the sum rule Eq.\ (\ref{sumrulesiam}).
The quantity gives the fraction of the adatom spin that remains unscreened by the substrate spins up to distance $r$ from $i_{1}$.
Using a 90\%-screening criterion, for example, i.e.\ $\varTheta(\xi_{\rm K}) = 0.1$, the extent of the cloud amounts to 
$\xi_{\rm K} \simeq 10-15$ lattice sites.
This is consistent with the discussion given above.
     
A criterion based on Eq.\ (\ref{cloud}) cannot precisely define the parameter range in which R-DMFT gives reliable results.        
Fig.\ \ref{local_substrate_chi} demonstrates that, using R-DMFT, the deviation from the numerically exact DMRG data grows {\em gradually} when decreasing the distance between the adatoms $d$.

\subsection{Distance dependence}

To estimate the reliability of the mean-field approach, we focus on the susceptibility $\chi_{i 1}^{\rm sub}$ at the substrate site below the first adatom $i=i_{1}$ where, according to the results shown in Fig.\ \ref{Substrate_chi}, the deviations are the strongest.
$\chi_{i_1 1}^{\rm sub}$ is shown in Fig.~\ref{local_substrate_chi} as a function of the distance $d$.
There is a nice quantitative agreement of the R-DMFT with the exact DMRG result for large $d$.
For smaller $d$, R-DMFT still predicts the correct trend, except for $d=1$.

\begin{figure}[t]
\includegraphics[width=0.4\textwidth]{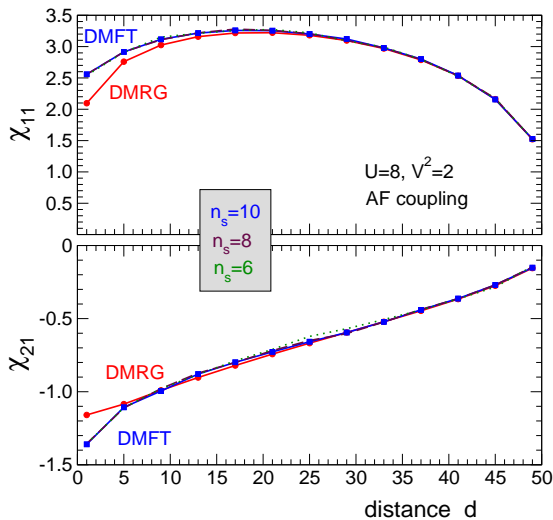}  
\caption{(Color online) Local adatom susceptibility $\chi_{11}$ and nonlocal adatom-adatom susceptibility $\chi_{21}$ as functions of the distance $d$ between the adatoms for $U=8$, $V=\sqrt{2}$ and for a system with $L=50$ substrate sites as obtained by R-DMFT and DMRG (red lines). 
R-DMFT calculations are done with different numbers of bath orbitals in the effective single-impurity models: $n_s=6,8,10$, as indicated.
}
\label{DMFT_vs_DMRG_d}
\end{figure}

For the same set of parameters Fig.\ \ref{DMFT_vs_DMRG_d} shows the local adatom susceptibility $\chi_{11}$ and nonlocal adatom-adatom susceptibility $\chi_{21}$ as functions of the distance $d=4n+1$ with integer $n$. 
In both cases the agreement of the R-DMFT with the DMRG results is excellent.
Significant differences are found for $d=1$ only and rapidly diminish with increasing $d$.

Fig.\ \ref{DMFT_vs_DMRG_d} includes R-DMFT results obtained with different $n_{s}$.
On the scale of the figure, there is no difference between the results for $n_{s}=8$ and $n_{s}=10$ bath sites in the effective single-impurity model while the results obtained for $\chi_{21}$ with $n_{s}=6$ slightly deviate for intermediate distances around $d=25$.
This comparison shows that the R-DMFT results are converged with respect to $n_{s}$. 
The differences to the DMRG data are thus intrinsic to the dynamical mean-field approach itself and not at all caused by discretization errors of the Lanczos solver. 

It is worth to mention that the distance dependence of $\chi_{21}$ cannot be explained by conventional RKKY theory.
For $J\to 0$, the magnetic susceptibility is determined by the effective two-spin Heisenberg model Eq.\ (\ref{eq:rkky}) which yields $\chi_{21}=-\chi_{11} \sim 1/J_{\rm RKKY}$ with $J_{\rm RKKY} \propto (-1)^{d}/d = 1/d$ at odd distances $d$. 
The decreasing absolute magnitude of $\chi_{21}$ with increasing $d$ and also the fact $\chi_{11}+\chi_{21}\not=0$ just reminds us that with $U=8$ and $V^{2}=2$ the system is well beyond the perturbative-in-$J$ regime and that there is a strong substrate contribution $\sum_i \chi_{i1}^{\rm sub}$ necessary to fulfill the sum rule Eq.\ (\ref{sumrulechi}).

For large $d$ the trends can rather be understood in a picture of two independent Kondo effects.
Clearly, $|\chi_{21}|$ is expected to decrease with $d$.
More interesting is the behavior of $\chi_{11}$ which develops a maximum around $d=15$--$20$.
The increase of $\chi_{11}$ with $d$ at short $d$ results from a reminiscence to the RKKY limit:
With increasing $d$ the effective coupling between the adatom decreases and their magnetic moments tend to become free resulting in a higher local susceptibility $\chi_{11}$.
Substrate contributions are sizable but cannot outweigh this effect in $\chi_{11}$, in contrast to $\chi_{21}$ which is 2--3 times smaller in absolute magnitude.
In the large-$d$ limit, where adatom-adatom interactions can be disregarded completely, one would naively expect a saturation of the local susceptibility at the inverse Kondo temperature since $\chi_{11} \propto 1/T_{\rm K}$ in a single-impurity model. \cite{Kondo}
However, $\chi_{11}$ must decrease since with increasing $d$ at fixed $L=50$ the adatoms move to the chain edges where we have a site-dependent Kondo temperature.
This increases with decreasing distance to the edge as the non-interacting substrate local density of states at the Fermi energy is increasing.

\begin{figure}[t]
\includegraphics[width=0.49\textwidth]{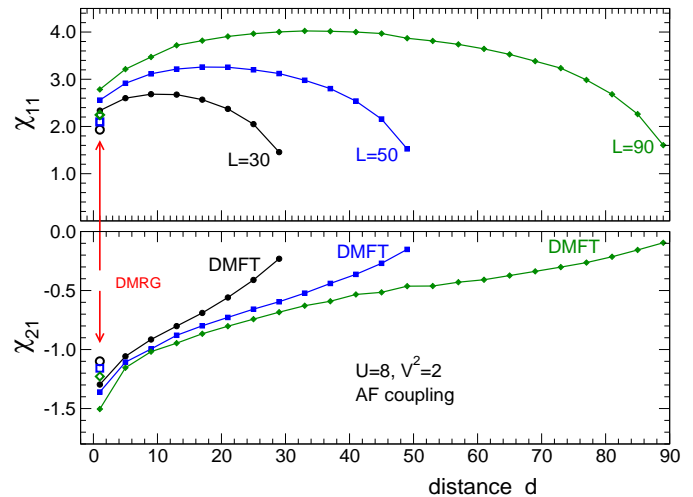}  
\caption{(Color online) Local and nonlocal adatom susceptibilities $\chi_{11}$ and $\chi_{21}$ as functions of distance $d$ for $U=8$, $V=\sqrt{2}$.
R-DMFT calculations for different system sizes $L$ as indicated.
DMRG data are shown for comparison at nearest-neighbor distance $d=1$ only.
}
\label{DMFT_L_d}
\end{figure}

This interpretation is corroborated by Fig.\ \ref{DMFT_L_d} which displays R-DMFT results for $\chi_{11}$ and $\chi_{21}$ for different system sizes $L=30$, $L=50$ and $L=90$.
We find the same qualitative behavior in all three cases.
Quantitatively, however, there are sizable differences at inter-adatom distance $d=1$, for example, which show that even with $L=90$ substrate sites the chain center cannot be regarded as bulk-like and that the center local density of states is still considerably dependent on $L$.
On the other hand, the susceptibilities for $d$ close to $L$, i.e.\ for systems with adatoms located at or very close to the chain edges, are almost converged.
Note that $\chi_{11}$ for $d=L$ is almost the same for $L=50$ and $L=90$.
Again this shows that, at least for the larger systems, the magnetic response is dictated by the physics of the single-site Kondo effect, i.e.\ the presence of the second adatom has almost no effect on $\chi_{11}$ and on the Kondo temperature of the first adatom. 
This does not exclude a finite magnetic interaction between the adatoms and in fact a non-zero $\chi_{21}$ for $d=L$ is found which, in addition, also does not depend on $L$ for the larger systems.

These adatom-adatom magnetic interactions are correctly captured by the R-DMFT.
R-DMFT and DMRG results coincide on the scale of the plot except for $d=1$ (DMRG results are shown for $d=1$ only).
This almost perfect agreement can be understood by referring to the strong-coupling limit:
For $J\to \infty$, the two adatom magnetic moments form perfectly local Kondo singlets that do not interact with each other. 
This limit is trivially accessible by the mean-field approach.
For finite coupling, second-order perturbation theory in $t/J$ predicts spin-spin correlations to decay as $1/d^2$. \cite{SA96,HE97} 
The inter-adatom magnetic interaction is thus expected to scale as $\propto J^{-4}$. 
This is also accessible to the R-DMFT approach while the neglected feedback of this effective interaction on local physical properties at one adatom, e.g.\ on $T_{K}$ and thus on $\chi_{11}$, is of higher order and small in the strong-$J$ limit.

\begin{figure}[t]
\includegraphics[width=0.49\textwidth]{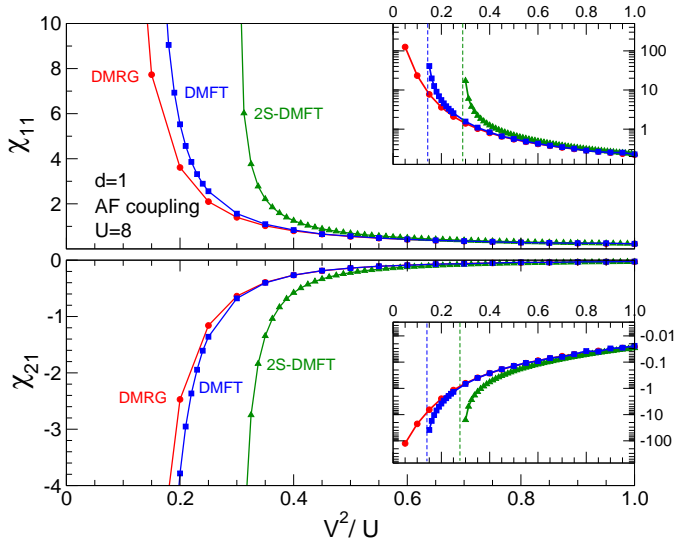}  
\caption{(Color online) Local and nonlocal adatom susceptibilities $\chi_{11}$ and $\chi_{21}$ as functions of $V^2/U$ for $U=8$ and $d=1$ as obtained by R-DMFT, real-space two-site DMFT and DMRG for a system with $L=50$ substrate sites. 
Insets: same quantities plotted on a logarithmic scale. 
Dashed lines indicate the critical $V^2/U$ where $\chi_{11}$ and $\chi_{21}$ diverge.
}
\label{DMFT_vs_2S-DMFT_DMRG_J}
\end{figure}

\begin{figure}[t]
\includegraphics[width=0.49\textwidth]{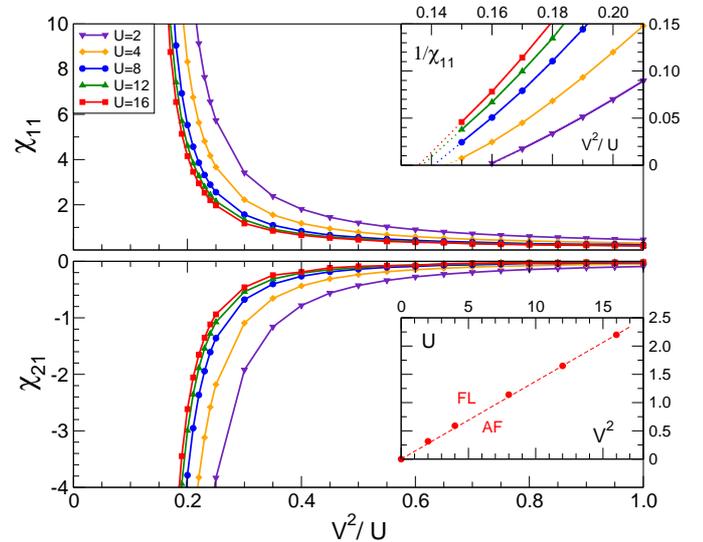}  
\caption{(Color online) Local and nonlocal adatom susceptibilities $\chi_{11}$ and $\chi_{21}$ as functions of $V^2/U$ for  $d=1$ and $L=50$ as obtained by R-DMFT. 
Results for different $U$ as indicated. 
Upper inset: inverse susceptibility $1/\chi_{11}$ and extrapolation (dotted lines) to $1/\chi_{11}=0$.
Lower inset: ``phase diagram'', separating the Fermi liquid state (FL) from an (artificial) antiferromagnetic state (AF) which shows up for weak $V$ and strong $U$.
}
\label{DMFT_U_J_PT}
\end{figure}

\subsection{Dependence on the local exchange coupling}

The breakdown of R-DMFT can be enforced, however, by decreasing $J$.
Fig.\ \ref{DMFT_vs_2S-DMFT_DMRG_J} shows the susceptibilities for $d=1$ as a function of $V^{2}/U$.
There is again excellent agreement for strong $V^{2}/U$ even with the simplified two-site R-DMFT. 
Deviations of the two-site approach from the exact $\chi_{11}$ and $\chi_{21}$ become sizable for couplings smaller than $V^{2}/U \approx 0.5$.
The full R-DMFT is reliable down to smaller values for $V^{2}/U$ but finally also starts to significantly deviate from the DMRG data for $V^{2}/U \lesssim 0.2$.
Here, as compared to the strong-coupling limit, the local susceptibility is by more than an order of magnitude higher, i.e.\ the Kondo temperature is by more than an order of magnitude smaller (see the upper inset).

For even smaller couplings, the mean-field approach breaks down completely and fails to maintain a Fermi-liquid ground state:
The small-$J$ limit is problematic for R-DMFT as the screening of the magnetic moments is too weak to compensate the ordering tendencies induced by a comparatively strong inter-adatom interaction. 
The system becomes too susceptible to an artificial spontaneous symmetry breaking that is induced by the mean-field approximation itself.
While the adatoms' state is given by a nonlocal SU(2) invariant singlet $(|\!\uparrow \downarrow \rangle - |\!\downarrow \uparrow \rangle ) / \sqrt{2}$ for $J\to 0$, the mean-field theory predicts an incoherent mixture of degenerate ordered states $|\!\uparrow \downarrow \rangle$ and $|\!\downarrow \uparrow \rangle$.

This qualitative failure is indicated by divergencies of $\chi_{11}$ and $\chi_{21}$ which take place at coupling strengths $V^{2}/U$ that are somewhat smaller than those where first quantitative deviations from the exact data were found (see insets in Fig.\ \ref{DMFT_vs_2S-DMFT_DMRG_J}).
This also implies that the mean-field approach is able to exhibit its limitations by itself.

Fig.\ \ref{DMFT_U_J_PT} shows the susceptibilities as obtained by R-DMFT for $d=1$ and different $U$ and $V$ as functions of $V^2/U$.
We find $\chi_{11}$ and $\chi_{21}$ to diverge at the same point in parameter space. 
From extrapolations of the inverse local susceptibility to $1/\chi_{11} = 0$ at different $U$ and $V$, shown in the upper inset, one may derive a mean-field ``phase diagram''. 
This is displayed in the lower inset. 
A normal Fermi-liquid ground state found for large $V$ and small $U$ is separated from the SU(2)-symmetry-broken antiferromagnetic state realized for small $V^2/U$.
This ``phase-transition line'' should actually be interpreted as a crossover from the Kondo to the RKKY regime or, more precisely, as the boundary up to which R-DMFT is reliable.
While the critical coupling $V^2/U$ is almost independent of $U$, there is some dependence on the distance $d$.
However, this is weak: 
While a Fermi-liquid ground state is obtained down to $V^2/U \approx 0.14$ for $d=1$, we find a slightly smaller critical value of $V^2/U \approx 0.12$ for $d=5$ and $V^2/U\approx 0.09$ for $d=49$ (at $U=8$).

\subsection{Different distances between the adatoms}
\label{sec:diffdist}

More important for the reliability of R-DMFT is the local Kondo temperature. 
This becomes obvious if the two adatoms are placed at a distance $d=4n+3$ with integer $n$, i.e.\ $d=3,7,11,...$ etc.
At the corresponding substrate sites $i_1$ and $i_2$ (symmetric to the chain center) we have a low weight $|U_{i_\alpha \ff k_{\rm F}}|^2$ of the one-particle energy eigenstate of the non-interacting substrate at the Fermi wave vector $\ff k_{\rm F}$ while $|U_{i_\alpha \ff k_{\rm F}}|^2$ is high for distances $d=4n+1$. 
This pronounced odd-even effect is a consequence of surface Friedel oscillations.
The weight $|U_{i_\alpha \ff k_{\rm F}}|^2$ determines the local substrate density of states and thus also the local Kondo temperature.
Consequently, $T_K$ is small for $d=4n+3$ and the nonlocal RKKY interaction much more efficient. 
Using DMRG, we in fact find $|\chi_{21}|$ at $d=3$ to be more than an order of magnitude larger than at $d=1$.
This effect even increases with increasing $d=4n+3$ since $|U_{i_\alpha \ff k_{\rm F}}|^2$ is decreasing if the adatoms move towards the chain edges.
At the edges ($d=47$) $|U_{i_\alpha \ff k_{\rm F}}|^2$ is suppressed by more than a factor 100 compared to the $d=4n+1$ case ($d=49$), and the Kondo temperature is essentially vanishing.
This regime is not accessible to R-DMFT.
While the mean-field approach predicts the correct sign, the absolute value $|\chi_{21}|$ and also $\chi_{11}$ is strongly underestimated for $d=4n+3$.
Deviations from the DMRG results grow with increasing $d$.
This had to be expected, as a huge nonlocal susceptibility $|\chi_{21}|$ induces via the Schwinger-Dyson equation a sizable contribution to the nonlocal self-energy which is neglected in R-DMFT.

The ferromagnetic case is different. 
Here we consider distances $d=2n$ with integer $n$.
In the Kondo limit of the model and for weak $J$, the ferromagnetic RKKY coupling of well-formed spins $1/2$ leads to a triplet ground state as is easily verified by means of DMRG calculations for $L=50$, i.e.\  there is a nonlocal spin $S=1$. 
In all our calculations this spin is not screened by the substrate electrons. 
This may be explained by the fact that $L=50$ is still too small to accommodate the corresponding screening cloud.
However, there is a nonlocal spin $S=1$ not only in the weak-$J$ limit. 
In fact, we find a triplet ground state for any choice of $U>0$ and $V\ne 0$.

Our R-DMFT calculations reproduce the spin-triplet ground state for small $V^2/U$ by predicting, for an infinitesimally small external magnetic field in $+z$ direction, a spontaneously symmetry-broken ferromagnetic state $|\! \uparrow \uparrow \rangle$.
This corresponds to the $M=1$ state of the DMRG spin triplet.
With increasing $V$, however, the expectation value of the $z$-component of the total spin $\ff S^{\rm tot} = \ff S^{f}_{1} + \ff S^{f}_{2} + \ff S^{\rm sub}$ deviates from unity and, beyond a critical hybridization $V$, even vanishes: $\langle S^{\rm tot}_z \rangle = 0$.
Hence, in the ferromagnetic case, R-DMFT is reliable in the small-$J$ but appears to fail in the strong-coupling limit.
This requires further investigations which, however, are beyond the scope of the present paper.

\section{CONCLUSION}
\label{con}

Conventional (RKKY) theory of indirect magnetic exchange predicts an effective exchange interaction $J_{\rm RKKY, ij} = J^{2} \chi_{ij}^{0,\rm sub}(\omega=0)$ where $\chi_{ij}(\omega=0)$ is the nonlocal static susceptibility of the metallic host.
This interaction survives, as a nearest-neighbor coupling, \cite{PP07} even in the case of two magnetic impurities embedded in an infinite-dimensional lattice and is thus accessible by dynamical mean-field theory.
In the limit of infinite spatial dimensions or, at finite dimensions, within the dynamical mean-field approximation, one can therefore expect a finite response at one magnetic impurity subject to a local magnetic field at the other one, located at nearest-neighbor but also for larger distances.

On the other hand, nonlocal effective interactions do not contribute to the single-particle self-energy on the DMFT level: The DMFT self-energy is just defined as the sum of the {\em local} skeleton diagrams only. 
This is a well-known shortcoming of mean-field theory which gives rise to artifacts in the RKKY limit.
Namely, for $J \to 0$ the magnetic impurities are only weakly coupled to the host and thus become extremely susceptible.
A tiny Weiss field within DMFT is then sufficient to drive the system to an artificial symmetry-broken state, i.e.\ an antiferromagnetic state rather than a nonlocal singlet of the impurity magnetic moments is formed.
A state with the characteristic distance dependence of the RKKY interaction, e.g.\ $J_{\rm RKKY} \propto 1/d$ for a one-dimensional system at half-filling, cannot be recovered within DMFT as it is always preempted by spontaneous symmetry breaking.
Clearly, solutions with a finite magnetic moment could easily be suppressed in a mean-field approach.
One should note, however, that the resulting magnetic susceptibility is unphysical, i.e.\ negative, as it refers to a thermodynamically unstable state.
Therefore, in any case, the physics of the RKKY limit is not accessible by DMFT.

The present study has shown, however, that beyond the RKKY limit, (real-space) DMFT is well suited to study even quantitatively the effects of indirect magnetic exchange.
Here, we have concentrated on two magnetic ``adatoms'' on a one-dimensional ``substrate surface'' -- a minimal model to study indirect magnetic interactions in competition with the Kondo effect for magnetic atoms on metallic surfaces and a model that is amenable to an exact numerical solution by means of the density-matrix renormalization group.

DMRG has been used to compute spin-spin correlation functions and static spin susceptibilities, and particularly adatom-substrate site correlations and susceptibilities.
Depending on the distance $d$ between the two adatoms and depending on the hybridization strength $V$, rather complicated profiles are obtained.
Comparing the results for the two-adatom (two-impurity) Anderson model with those obtained for the corresponding single-adatom (single-impurity) Anderson model, one can easily classify the different features of those profiles as single-impurity effects or as resulting from the adatom-adatom effective interaction.
In this way, clear reminiscences of the RKKY interaction, i.e.\ of nonlocal singlet formation, are found to compete with the formation of Kondo clouds and screening of the adatom magnetic moments.
In addition, the profiles are strongly affected by the finite system size (chains with typically $L=50$ have been considered here) and by effects resulting from strong surface Friedel oscillations in the local density of states, especially if the adatoms are in the vicinity to one of the chain edges.
This complex interplay of different physical mechanisms is almost perfectly recovered by the real-space DMFT.

Qualitatively, the real-space DMFT is reliable as long as the model parameters, in particular the local exchange coupling $J \propto V^2/U$, are in a regime well separated from the artificial symmetry-broken state.
This parameter regime, where the adatom susceptibilities are not too large or where the adatom magnetic moments are predominantly interacting with the substrate moments rather than among each other, however, goes well beyond the extreme Kondo limit of non-overlapping Kondo clouds.
The critical value for $V^2/U \approx 0.14$ in units of $t$ at $d=1$ gives an impression of a lower bound for the applicability of R-DMFT.

While the present study has focused on a one-dimensional model to allow for benchmarking against numerically exact DMRG results, future applications of the R-DMFT should address higher-dimensional systems. 
With increasing coordination number of correlated sites, the parameter space accessible to the mean-field approach is expected to be become larger or mean-field artifacts less pronounced.
For the case of atoms trapped in optical lattices, there are impressing examples where R-DMFT has contributed to an understanding of the physics of inhomogeneous systems with ${\cal O}(100)$ correlated and geometrically inequivalent sites in two dimensions, for example. \cite{RDMFT-ultracold}

Let us also point out that for Anderson-type multi-impurity or lattice models, one typically expects non-local magnetic correlations to diminish rapidly as the electron filling on the correlated sites is changed away from half-filling. 
Systems off half-filling are thus expected to be more amenable to an R-DMFT approach.
At the same time they are also interesting physically as reducing the filling away from half-filling affects local-moment formation as well. 
Hence, the competition between non-local RKKY interaction and Kondo screening must be seen as strongly filling dependent.

For complex magnetic nanostructures with several magnetic adatoms in different chain or cluster geometries on two- or on semi-infinite three-dimensional metallic surfaces, a mean-field approach is inevitable anyway.
Here the conceptual simplicity of a single-site mean-field theory, as compared to different possibilities for cluster extensions, is important as it allows to study almost arbitrary geometries. 
As in ab-initio studies, the accessible system size strongly depends on the remaining e.g.\ lateral spatial symmetries, and the computational effort scales nearly linearly with the number of inequivalent correlated sites only.

The two-impurity one-dimensional Anderson model represents a model that is rather unfavorable to a single-site R-DMFT approach. 
Even for this case, as the present study has shown, R-DMFT can in fact almost quantitatively predict the effects of indirect magnetic exchange in competition with the Kondo and with geometrical effects -- as long as the approximation predicts a Fermi-liquid ground state.

\section*{Acknowledgments}

The work was supported by the Deutsche Forschungsgemeinschaft within the Sonderforschungsbereich 668 (project A14).
N.R. acknowledges support by the Forschergruppe FOR 1346.


\end{document}